# Assessment of the application of the Universal Competencies

S N Masaev[1,7,8], G A Dorrer[1,2], A N Minkin[1,3], A V Bogdanov[1,4] and Y K Salal[5,6]

[1] Siberian Federal University, pr. Svobodnyj, 79, Krasnoyarsk, 660041, Russia
[2] Department of System Analysis and Operations Research, Office L-409, 410, 31, Krasnoyarsky Rabochy Av., Reshetnev Siberian State University of Science and Technology, Krasnoyarsk, 660037, Russia
[3] FSBEI HE Siberian Fire and Rescue Academy EMERCOM of Russia, 1 Severnaya Street, Zheleznogorsk, 662972, Russia
[4] The Main Directorate of EMERCOM of Russia for Krasnoyarsk Territory, Mira Ave., 68, Krasnoyarsk, 660122, Russia
[5] South Ural State University, pr. Lenina 76, Chelyabinsk, 454080, Russia
[6] University of Al-Qadisiyah, Diwaniyah, 58002, Iraq
[7] Control Systems LLC, 86 Pavlova Street, Krasnoyarsk, 660122, Russia

[8] E-mail: faberi@list.ru

**Abstract**. Application of Universal Competencies in Russian educational institutions is very important. Based on them, educational standards are invented. However, there is no universal assessment of the application of the Universal Competencies in practice. The main idea of the research is a general assessment of the application of universal competencies. For this, the activity of the enterprise is modeled. The enterprise process model is combined with the Universal Competencies. Further, the measurement is made by a universal indicator. The analysis of the dynamics of the universal indicator proves the existence of an assessment of the application of the Universal Competencies at a production facility. The integral indicator is a universal assessment of the application of the Universal Competencies.

## 1. Introduction

The idea of creating common competencies is reaching more and more participants. Today it is the mainstream that has swept across Russia. Large educational organizations in Russia focus on Universal Competencies [1].

The idea of universal competencies arose when European countries were united. The unification of European countries resulted in the unification of industry. There was a request from enterprises for uniform requirements for the skills of workers employed in other European countries.

Universal competences were announced long ago, back in 1996 in Bern within the framework of the Council of Europe [1], but they did not receive further significant development. They include competencies: autonomous action (independence and individual initiative), the use of tools (physical and socio-cultural means, including a computer, natural language, etc.), functioning in socially heterogeneous groups (tolerance, willingness to interact with people). In addition, political and social competence (non-violent conflict resolution, maintenance of democratic institutions), a critical attitude to information in the media and advertising, the ability to learn throughout life, etc. are highlighted. [1]

Competencies acquired in universities can be represented in different ways. For example, interpret them through Universal Competencies (recommended by the Council of Europe) [1], Dublin Descriptors [2], European Qualifications Framework (passport of qualifications) [3], European Qualifications Framework for EU countries [4], National Qualifications Framework [5].

The above qualification documents are united by the creation of universal competencies within the Bologna process [4].

Various issues of competence research are reflected in the works of: V.P.Bespalko, N.V. Kuzmina, I. Ya. Lerner, A. K. Markova, G. V. Mukhametzyanova, A. M. Novikov, A. S. Simonov, V. A. Slastenin, S. N. Chistyakova, K. A. Abulkhanova-Slavskaya, B. G. Ananyeva, A. A. Bodalev, V. V. Davydova, L. S. Vygotsky, P. Ya. Galperin, I. A. Zimney, A. A. Leontyev, A. N. Leontyev, S. L. Rubinshtein, V. A. Bolotov, V. V. Kraevsky, O. E. Lebedeva, V. V. Serikova, M. A. Kholodnaya, A. V. Khutorsky, T. I. Shamova, A. A. Pinsky, I. D. Frumin, B. D. Elkonin, E. F. Zeer, V. S. Ledneva, R. P. Milruda and others.

The process of creating universal competencies generates a huge process of iterating competencies from other methodologies. There is a huge analysis of the applicability of each competency to various industries. After that, update the curricula of educational institutions.

Universal competencies allow you to develop more advanced training programs in educational organizations. Universal competencies allow graduates of educational institutions to more quickly adapt to work in enterprises. This will reduce the huge amount of work to improve the list of universal competencies.

Therefore, the purposegive a universal assessment of the application of universal competencies at a production facility.

We going to do a tasks for that:
- Create an enterprise model;
- Introduce Universal Competencies into the Enterprise Model;
- Assess the level of application of Universal competencies;
- Analysis of results.

To create a model of a production facility, models of classical works were used [6-8].

## 2. Method

Step 1. Many processes of the enterprise are formed $X$. Each $t$ time period is analyzed for the presence of an ongoing or non-running process. Then a system of processes is formed $S=\{T,X\}$, where $T=\{t:t=1,...,T_{max}\}$ - a lot of time points, $x(t)=[x^1(t),x^2(t),...,x^n(t)]^T \in X - n -$ vector of indicators process. Indicators of the vector $x^i(t)$ - the value of financial expenses and income of the enterprise [9,10]. $S=\{T,X\}$ identified as a model of the enterprise. Classical surveillance and control tasks apply to model [7-11].

Step 2. Comparison of Universal competencies $v(t)$ with processes $x(t)$ of the enterprise $S=\{T,X\}$. We form $v_i^j$ (compliance $x_j^i$ is $v_i^j$ set as 1-yes, 0-no) from $i$ – from competence of university graduates and $j$ process model. We get Universal competencies set $V$ where $v(t) = \left[v_1^1(t), v_2^j(t),...,v_m^n(t)\right]^T \in V$.

Payment competencies of the model is limited by resources $C$, then $C(V) \leq C$. This restriction applies to all subsystems of the researched system.

Step 3. Calculation of the integral $V$ index through the correlation matrix $R_i(x)$

$$V_i(t) = R_i(t) = \sum_{j=1}^{n} |r_{ij}(t)|. \tag{1}$$

$$R_k(t) = \frac{1}{k-1} V_k^{oT}(t) V_k^o(t) = \|r_{ij}(t)\|, \tag{2}$$

$$r_{ij}(t) = \frac{1}{k-1} \sum_{l=1}^{k} v^{i}{}^{o}(t-l) v^{j}{}^{o}(t-l), \quad i,j = 1,...,n, \tag{3}$$

$$V_{k}(t) = \begin{bmatrix} v^{T}(t-1) \\ v^{T}(t-2) \\ \cdots \\ v^{T}(t-k) \end{bmatrix} = \begin{bmatrix} v^{1}(t-k) & v^{2}(t-k) & \cdots & v^{n}(t-k) \\ v^{1}(t-k) & v^{2}(t-k) & \cdots & v^{n}(t-k) \\ \cdots & \cdots & \cdots & \cdots \\ v^{1}(t-k) & v^{2}(t-k) & \cdots & v^{n}(t-k) \end{bmatrix} \tag{4}$$

where $t$ are the time instants, $r_{ij}(t)$ are the correlation coefficients of the variables $v^i(t)$ and $v^j(t)$ at the time instant $t$.

Step 4. The analysis of experimental data is performed graphically. The dynamics of the integral indicator is calculated for all periods of time.

$$V = \sum_{t=1}^{T=\max} \sum_{i=1}^{n} V_i(t). \tag{5}$$

## 3. Characteristics of the research objects

Universal competencies in different specialties: 1. operate with legal regulations within the framework of laws related to professional, 2. search for opportunities for continuous self-development, 3. take responsibility for the results of activities (your own and other people), 4. identify problems and use adequate technologies to solve them, 5. gets a bottom line, 6. organize your own activities (practical, cognitive), 7. initiate performance improvement, 8. put forward innovative ideas and non-standard approaches to their implementation, 9 organize the activities of others, 10 focus on the consumer, 11. realize that you are a citizen of the country and take responsibility for your civil position, 12 lead projects, 13. perform functions, 14. builds interpersonal relationships, 15. build interpersonal interactions, 16. identify and manage conflicts, 17. carry out oral and written communication in native and foreign, 18. listen to interlocutors using different, 19. negotiate, 20. participate in the exchange of ideas, information and knowledge with others, 21. confidently defend your position when objections arise, 22. present plans and results of own and team activities using various means, 23. assess the consequences of the impact of professional activities on the environment, 24. carry out reflection in relation to yourself, 25. tolerantly perceive different cultures and be able to work with representatives of different cultures, 26. critically assess yourself and others, and take critical judgments constructively, 27. integrate knowledge from different fields to solve professional problems, 28. Have basic knowledge over a professional / interdisciplinary nature, 29. carry out actions to search, analyze, systematize and evaluate information, 30. use the principles of systemic, 31. use information technology for processing, presentation, transmission and storage, 32. create your own positive image [1].

The application of Universal competencies is being considered at a woodworking enterprise in the city of Krasnoyarsk, Krasnoyarsk Territory. The average headcount of the enterprise is 650 people.

Competencies in specialties versus enterprise model. The enterprise model is characterized by 1.2 million parameters. Simulation is performed in the software package [9].

## 4. Experiment result

Initial calculation data: $n$=1.2 million values, $X$=5,641,442 thousand rubles, control is set through Universal competencies ($V_{Universal\_competencies}$). From the 7th period, three managers and three personnel managers are hired in accordance with the directions of the enterprise's activities to maintain the selected management loop at the enterprise. From the 13th period, two managers are dismissed, since after the introduction of universal competencies, their functions are transferred to personnel managers. The calculation algorithm is 423 minutes [9].

A table 1 shows the experiment result of estimating the control mode $V_i(t)$ through Universal competencies.

**Table 1.** Regimes: $V_{(basic\ mode)}$ and $V_{(Universal\_Competencies)}$.

| t | $V_{(basic\_mode)}$ | $V_{(Universal\_Competencies)}$ | $\Delta V$ | t | $V_{(basic\_mode)}$ | $V_{(Universal\_Competencies)}$ | $\Delta V$ |
|---|---|---|---|---|---|---|---|
| 1 | 87.34 | 110.67 | 23.33 | 30 | 96.32 | 95.32 | -1.00 |
| 2 | 70.94 | 90.32 | 19.38 | 31 | 105.10 | 104.10 | -1.00 |
| 3 | 51.43 | 69.05 | 17.62 | 32 | 98.66 | 97.66 | -1.00 |
| 4 | 56.35 | 81.39 | 25.04 | 33 | 82.19 | 81.19 | -1.00 |
| 5 | 59.26 | 89.87 | 30.61 | 34 | 76.23 | 75.23 | -1.00 |
| 6 | 73.39 | 109.99 | 36.60 | 35 | 68.52 | 68.52 | 0.00 |
| 7 | 95.25 | 123.50 | 28.25 | 36 | 60.51 | 60.52 | 0.00 |
| 8 | 92.64 | 124.07 | 31.43 | 37 | 53.13 | 53.13 | 0.00 |
| 9 | 95.53 | 125.43 | 29.90 | 38 | 61.65 | 61.65 | 0.00 |
| 10 | 70.17 | 100.32 | 30.15 | 39 | 53.51 | 53.51 | 0.00 |
| 11 | 58.42 | 74.76 | 16.35 | 40 | 51.84 | 51.84 | 0.00 |
| 12 | 56.48 | 72.97 | 16.50 | 41 | 72.03 | 72.03 | 0.00 |
| 13 | 61.88 | 79.69 | 17.81 | 42 | 93.08 | 93.08 | 0.00 |
| 14 | 71.87 | 90.80 | 18.93 | 43 | 99.23 | 99.23 | 0.01 |
| 15 | 52.45 | 67.93 | 15.48 | 44 | 115.79 | 115.80 | 0.00 |
| 16 | 53.92 | 70.05 | 16.14 | 45 | 110.12 | 110.12 | 0.00 |
| 17 | 84.06 | 101.55 | 17.48 | 46 | 103.64 | 103.64 | 0.00 |
| 18 | 114.43 | 132.62 | 18.19 | 47 | 88.22 | 87.22 | -0.99 |
| 19 | 132.20 | 132.20 | 0.00 | 48 | 69.27 | 68.27 | -1.00 |
| 20 | 153.90 | 153.91 | 0.00 | 49 | 55.14 | 54.14 | -1.00 |
| 21 | 164.54 | 164.54 | 0.00 | 50 | 63.51 | 62.51 | -1.00 |
| 22 | 150.02 | 151.03 | 1.01 | 51 | 50.02 | 49.02 | -1.00 |
| 23 | 140.74 | 144.75 | 4.01 | 52 | 61.58 | 60.58 | -1.00 |
| 24 | 115.09 | 119.10 | 4.00 | 53 | 60.33 | 59.33 | -1.00 |
| 25 | 87.02 | 91.02 | 4.00 | 54 | 147.33 | 147.33 | 0.00 |
| 26 | 100.60 | 104.60 | 4.00 | 55 | 158.41 | 158.41 | 0.00 |
| 27 | 87.59 | 91.59 | 4.00 | 56 | 156.87 | 156.87 | 0.00 |
| 28 | 76.04 | 79.04 | 3.00 | 57 | 167.90 | 167.90 | 0.00 |
| 29 | 76.26 | 76.26 | 0.00 | **Total** | **5 069.93** | **5 491.17** | **421.24** |

Figure 1 shows the experiment result of estimating the control mode $V_i(t)$ through Universal competencies.

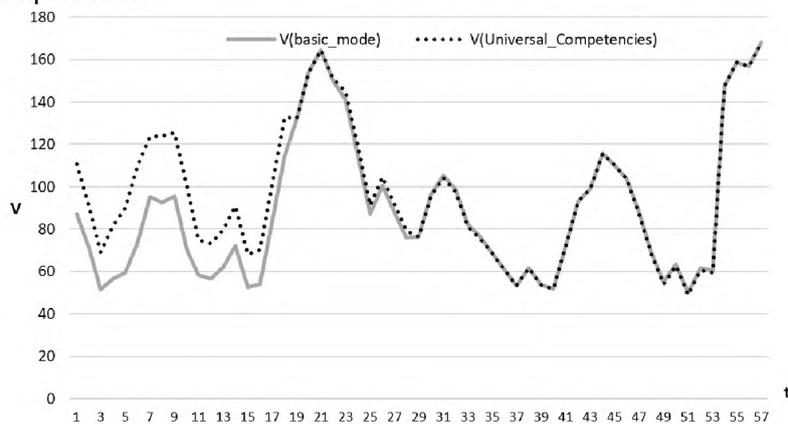

**Figure 1.** Indicator dynamics $V_i(t)$.

## 5. The discussion of the results

The dynamics of the integral indicator reflects the transition of an enterprise to management through Universal Competencies. This means that instead of the functional responsibilities of the staff, its competencies are considered.

The integral indicator (universal indicator) records the introduction of Universal skills into the system and their further use by the enterprise personnel. Personnel management through Universal Indicators is possible, as well as by students.

2,936 thousand rubles were spent on measuring the application of Universal competencies at the enterprise. Then the total costs of the enterprise for five years will amount to 5,644,378 thousand rubles. This means that the indicator records organizational measures for the implementation of the assessment of the application of Universal competencies at the enterprise.

Having a universal assessment of universal indicators, it is possible to analyze their relevance in industry. This will make it possible to determine exactly how much and what profession it is necessary to train workers. Universal competences will allow to combine the activities of various educational organizations in Russia and abroad. For example, the integration of education in the Countries of Independent States.

The universal assessment of the application of the Universal Competencies is a step towards these opportunities.

## 6. Conclusion

Having an assessment of universal competencies, they can be compared with other known approaches to the formation of competencies. Use universal competencies for the formation of personnel in special economic zones [11] and other tasks. Find contradictions or general patterns. The main thing is to improve them, etc.

The research tasks were completed:

- Created a model of the enterprise $S=\{T,X\}$;
- Created an enterprise model $v(t) = \left[ v_1^1(t), v_2^j(t), ..., v_m^n(t) \right]^T \in V$ into an enterprise model;
- The level of application of universal competences was assessed $V_{(Universal\_Competencies)}$;
- Results analysis performed $\Delta V = V_{(Universal\_Competencies)} - V_{(basic\_mode)} = 5,491.17 - 5,069.93 = 421.24$.

The purpose of the research has been achieved.